\documentclass[12pt]{article}
\usepackage{amsmath,amssymb,graphicx,mathrsfs,hyperref}

\newcommand{\be}{\begin{equation}}
\newcommand{\ee}{\end{equation}}
\newcommand{\bea}{\begin{eqnarray}}
\newcommand{\eea}{\end{eqnarray}}

\def\({\left(} \def\){\right)}

\begin{document}

\title{\vspace{-1.8in}
\vspace{0.3cm} {\Large Graviton n-point functions \\
for  UV-complete theories   in Anti-de Sitter space}}
\author{\large Ram Brustein${}^{(1,2)}$,  A.J.M. Medved${}^{(3)}$ \\
 \hspace{-1.5in} \vbox{
 \begin{flushleft}
  $^{\textrm{\normalsize
(1)\ Department of Physics, Ben-Gurion University,
    Beer-Sheva 84105, Israel}}$  $^{\textrm{\normalsize
 (2)  CAS, Ludwig-Maximilians-Universit\"at M\"unchen, 80333 M\"unchen, Germany}}$
$^{\textrm{\normalsize (3)  Department of Physics \& Electronics, Rhodes University,
  Grahamstown 6140, South Africa }}$
 \\ \small \hspace{1.7in}
    ramyb@bgu.ac.il,\  j.medved@ru.ac.za
\end{flushleft}
}}
\date{}
\maketitle

\begin{abstract}

We calculate graviton $n$-point functions in an anti-de Sitter black brane background for effective gravity theories whose linearized equations of motion have at most two time derivatives. We compare the $n$-point functions in Einstein gravity to those in theories whose leading correction is quadratic in the Riemann tensor. The comparison is made for any number of gravitons and for all physical graviton modes in a kinematic region for which the leading correction can significantly modify the Einstein result. We find that the $n$-point functions of Einstein gravity depend on at most a single angle, whereas those of the corrected theories may depend on two angles. For the four-point functions, Einstein gravity exhibits linear dependence on the Mandelstam variable $s$ versus a quadratic dependence on  $s$ for the corrected theory.
\end{abstract}
\newpage

\section{Introduction}

\subsection{Motivation and Objectives}

The gauge--gravity duality \cite{Maldacena1,Witten,Witten2,Maldacena} can be used  to relate
properties of a strongly coupled fluid to those of a  weakly coupled theory of anti-de Sitter (AdS) gravity \cite{hydrorev} (and references therein). A vast literature is devoted to using graviton and other two-point functions as a means for calculating the  two-point correlations of various operators in the gauge theory; see, however, \cite{Mald-Hof,Hof}.
In particular, the ratio of the shear viscosity to the entropy density $\eta/s$ has been a focal point of attention
\cite{PSS,KSS-hep-th/0309213,SS}.

The present treatment broadens the scope to graviton $n$-point functions for arbitrary $n$. This is meant as  preparation for using the corresponding multi-point correlation functions of the gauge-theory stress tensor as a probe of the gravitational dual of the quark--gluon plasma. This plasma  is produced in heavy-ion collisions and, so, of direct observational relevance \cite{QCD}.

The key idea is a recent observation \cite{prepap}
that the effective theory describing  gravitational perturbations
about a background solution  is,  itself, highly constrained
irrespective of the exact details of the UV-complete theory.
The argument is based on considerations of unitarity, which
follows naturally from the property of  UV-completeness
on both sides of the gauge--gravity correspondence.

The argument in \cite{prepap} is that we should only consider theories whose linearized equation of motion for the gravitons has, at most, two time derivatives. The non-linear interactions of such theories are constrained only by general covariance, which can be contrasted with Lovelock's original construction
\cite{LL}.
The latter further constrains the form of the interaction terms and limits them to a small finite number for each spacetime dimensionality. This implies that the effective theory of perturbations is of the ``Lovelock class" of gravitational models, as defined in detail below. This class contains the Einstein (two-derivative) and Gauss--Bonnet (four-derivative) terms, plus  a series of terms  with ever-increasing numbers of derivatives. In spite of the higher-derivative extensions, all Lovelock class theories satisfy, by construction, the two-derivative constraint on the
equation of motion \cite{LL}.

Part of the motivation for  the current work is the prospect of
an experimental  test of the multi-particle correlations in heavy-ion collisions.  The purpose is to initiate  this task, which is accomplished as follows:

We assume an AdS black brane background geometry and calculate the graviton $n$-point functions for both relevant theories.
Considerations are limited  to a kinematic regime
of a ``high momentum'',  which is defined  further on.
Otherwise, we determine all the physically relevant $n$-point functions for any number of gravitons.

The restriction to the high-momentum kinematic region is chosen with two reasons in mind. First, this kinematic region  allows for the suppressed Gauss--Bonnet corrections to compete in the best way with the leading-order Einstein results. Second, this region manifestly reveals how the two theories are
fundamentally distinct: Because Einstein gravity
is polarization independent, its  $n$-point functions depend on at most a single scattering angle, whereas the Gauss--Bonnet theory is polarization dependent and its $n$-point functions typically depend on two angles. For the 4-point functions, the distinction is expressed through a quadratic dependence on the Mandelstam variable $s$ for
the Gauss--Bonnet theory  versus a linear dependence on $s$ for Einstein's.

\subsection{The meaning of ``Lovelock class'' theories}

We wish to explain in more detail
how the condition of having at most two time derivatives in the linearized equations of motion limits the possible class of gravity theories. As is well known and will be evident
from Subsection~4.2, the only term  in
the Gauss--Bonnet Lagrangian which is physically significant
is the Riemann-squared term. The other two terms are, essentially, ``along for the ride''  so as to assure that the  equations of motion contain no more than two (time) derivatives. But even this statement can be deceiving, as the two extra terms can be viewed as an artifact of a particular choice of metric variables \cite{tH,POL} (and, again, \S 4.2). And so it is more accurate
to say: ``the four-derivative unitary extension of Einstein gravity is defined by adding a Riemann-tensor-squared term  to the Einstein-Hilbert action, supplemented by boundary conditions that ensure a two-derivative (linearized) field equation.'' This, of course, implies that sources for the exorcized modes are not allowed either.

A similar statement should apply to a unitary extension of
Einstein gravity to arbitrary order in the number of derivatives.
For this reason, six- and higher-derivative corrections can still play a role in a five-dimensional spacetime despite the fact that the Lovelock series
terminates at the  Gauss--Bonnet
extension.~\footnote{This point was missed by us
previously.} However, such corrections are suppressed  by factors of momentum divided by the cutoff scale of the gravity theory or, equivalently, by inverse powers of the 't Hooft coupling of the gauge theory. For this reason, we will
limit the current considerations to Einstein gravity and its
leading-order (four-derivative) correction.

Nevertheless, we will, to avoid confusion and clutter, continue to adhere with the standard nomenclature such as Gauss-Bonnet, Lovelock, {\it etcetera}.

\subsection{Difference between Einstein and Lovelock theories}

The extra pair of derivatives  of  Gauss--Bonnet gravity
is  directly responsible for one of its two
physical distinctions  with Einstein's
theory; namely, the structure of the higher- (than two)
point functions. The other physical difference
is that Gauss-Bonnet theories disobey Einstein's  equivalence principle. This first distinction is essential for the following reason: The defining feature of any Lovelock theory is that the linearized field equation is at most quadratic in derivatives. Hence, the two-point functions of Lovelock tend to  all  look rather the same, at least when compared at a fixed choice of polarization. Conversely, the higher-point functions can and will
be substantially different.

This difference in the higher-point functions becomes apparent when these are re-expressed in terms of scattering angles. As will be made clear, Einstein is unique among gravity theories in that any of its $n$-point functions depend on at most a single independent angle. This outcome can  be viewed as a consequence of a redundancy that  was already alluded to by Hofman \cite{Hof}. He demonstrated that, for a strictly two-derivative theory, the higher-point functions carry what is redundant information about the propagator. We see this quite literally in the
current work, inasmuch as any Einstein $2n$-point function could be obtained directly from a two-point function, using only simple combinatorial arguments.

The simple nature of the  Einstein angular dependence,  when compared to Gauss-Bonnet
and other higher-derivative theories, is already well understood from the work of Hofman and Maldacena \cite{Mald-Hof}.  There, however, the more complicated angular dependence of such  ``non-Einstein'' models  is viewed from the field-theory perspective and attributed  to a discrepancy in  the central charges of the gauge-theory dual.  This discrepancy is absent  for gauge theories with Einstein duals but generally is not.
Our main  point here is that this distinction could already be deduced from the bulk point of view without detailed knowledge about the gauge theory.

As an aside, let us point out that the same logic that underlies this distinction between Einstein and Gauss--Bonnet gravity can be extended to the purpose of comparing  Lovelock models of arbitrary order.  Just as the four- and higher-point functions are redundant for Einstein, the same must  be true for the six- and higher-point functions of Gauss--Bonnet. Then, with each additional inclusion of a term from the Lovelock series, the order that this redundancy sets in will increase accordingly. So that, in scenarios where higher-order Lovelock extensions could be relevant, there is an in-principle means of distinguishing the different  models by looking at $2n$-point functions with  an  increasingly larger value of  $n$.

\subsection{Contents}

The rest of the paper proceeds as follows.
The next section describes the basic set-up and strategies,
introduces some important formulas and
fixes conventions.   Sections~3 and~4  are dedicated to
calculating the graviton $2n$-point
functions (a function with an odd number of gravitons vanishes trivially) for the Einstein and Gauss--Bonnet cases, respectively. In Section~5, we elaborate on and  substantiate  the statements about angular dependence, as well as  make the connection to the gauge theory. Section~6 summarizes our conclusions.
One of the  supporting calculations is  deferred to an appendix.

\section{The basic framework}

Our starting point is a gauge field theory and its
presumed AdS gravitational dual.
The premise is to learn about  the strongly coupled properties of the former
from the weakly coupled limit of the latter.
We assume that the AdS bulk spacetime  is described by
string theory, which is a  unitary and UV-complete theory.
It is also assumed that the full UV completion
can be approximated by a gravitational action
that includes the Einstein  term along with
higher-derivative corrections.
Lastly, we assume that the action's  equations of motion
support a stationary black brane solution, as this geometry
will serve  as the background.

\subsection{Formalism and conventions}

A  $D$-dimensional (asymptotically) AdS  black brane can  be described by the following background metric:
\be
ds^2\;=\;-f(r)dt^2+ \frac{dr^2}{g(r)}+\frac{r^2}{L^2} dx_i^2\;.
\label{geometric}
\ee
The index denotes the transverse space dimensions $i=1,\dots,D-2$, $L$ is the AdS radius of curvature and $r$ is the radial coordinate
(orthogonal to the brane).
The functions $f$ and $g$ are constrained to
asymptote to $r^2/L^2$
at the AdS boundary
($r\to\infty$)
and  vanish on the horizon  ($r=r_h$); meaning that all
AdS brane solutions look
exactly the same on these two surfaces (the latter because
of ``no-hair'' theorems).

Irrespective of the higher-derivative terms
and other matter fields in the string theory, we can expect
$f$ and $g$ to agree with their Einstein forms,
 $f, g = \;\frac{r^2}{L^2}\left(1-\frac{r_h^{D-1}}
{r^{D-1}}\right)\;$, up to perturbatively
small corrections. These corrections are, on general grounds,
of  the order $\;l^2_p/L^2\;$ with $l_P$ being the Planck length.  From now on, we set $L=1$ unless stated otherwise.

Small metric perturbations  about this background  solution,
$\;g_{ab}\to g_{ab}+h_{ab}\;$,
should have a description  in terms of an effective field theory.
The effective model will naturally inherit
higher-derivative corrections; however, as explained in \cite{prepap} and commented above,
unitarity constrains  these corrections to be organized into
Lovelock extensions of Einstein's theory.

The immediate aim is  to calculate the graviton $n$-point  functions for arbitrary $n$. As discussed in \cite{BGH-0712.3206}, these functions can be viewed as a measure of the
gravitational coupling between $n$ interacting gravitons
and, so,  can be determined by expanding out
the Lagrangian density $\sqrt{-g}{\cal L}$ to the relevant
perturbative order. To this end, it is useful to define the tensor
\be
{\cal X}^{abcd}\;\equiv\; \frac{\partial {\cal L}}{\partial {\cal R}_{abcd}}\;.
\ee
For later use,   ${\cal X}^{abcd}$ inherits all of the
(anti-) symmetry properties of the Riemann tensor ${\cal R}^{abcd}$
and, for Lovelock theories  in particular, must satisfy the
identity \cite{LL}
\be
\nabla_a{\cal X}^{abcd}\;=\;0\;.
\label{LLidentity}
\ee

As is standard procedure in the analysis of AdS brane models, we impose the radial gauge on the gravitons or
$\;h_{ra}=0\;$ for any choice of $a$.
This gauge allows us to separate the gravitons into
three sectors: tensor, vector and scalar, which are sometimes also called transverse/traceless, shear and sound \cite{PSS2}.
With $z$ denoting the direction of graviton propagation parallel to the brane and $x$, $y$,  any pair of transverse brane directions that are orthogonal to $z$, these sectors can be classified respectively as
\be
\;h_2\;=\;\left\{h_{xy}\right\}\;
\ee
\be
\;h_1\;=\;\left\{h_{zx},h_{tx}\right\}\;
\ee
\be
\;h_0\;=\;\left\{h_{tt},h_{zz},h_{zt},h_{\;\;x_{i}}^{x_i}\right\}\;.
\ee

Only special combinations of the modes have physical gauge-invariant meaning \cite{KSonly}. Respectively, these are
\begin{eqnarray}
h_{xy}\;,\;\;\;\;\; \nabla_t h_{zx}-\nabla_z h_{tx}\;,\;\;\;\;\;
\Box h^a_{\;\;a}\;. \nonumber
\end{eqnarray}

The scalar-mode interactions involve at least two derivatives and, as explained
in Appendix~\ref{argue}, any occurrence of a $\nabla_a\nabla_b h_{cd}$ can be eliminated via the equations of motion. Hence, the scalar mode decouples on-shell.
A physical scalar mode requires an external source in addition to the background brane.

Thus, we are left to consider the tensor and vector modes.
Vector interactions involve at least one derivative per mode, and so their maximum number in a  $2n$-point function
is set by  the highest derivative term of the gravity theory.
To understand why, let us consider  the coordinate transformation
$\;x^{a}\to x^{a}+\xi^{x_i}\delta^a_{x_i}\;$  such that $\;\nabla_t \xi_{x_i}
= -h_{tx_{i}}\;$.
Then  $\;h_{tx_{i}}$ is set to $0\;$ but, as readily verified,
$\;\nabla_z h_{tx} - \nabla_t h_{zx}\;$  does not change.

Physically, this can be understood by the vector modes having an
effective description as components of an electromagnetic vector potential \cite{KSS-hep-th/0309213}.  For instance, the compactification of  $x$ reduces $D=5$ Einstein gravity to a $D=4$ Einstein--Maxwell theory such that
$\;A_0=h_{tx}\;$ and $\;A_z=h_{zx}\;$.
We will, therefore,  sometimes
write the vector  modes as
\be
F^{(j)}_{tz}\;\equiv\; \omega_j h^{(j)}_{zx}+ k_j h^{(j)}_{tx}\;.
\label{FST}
\ee
The choice of notation emphasizes that, as far as these modes are concerned,
the
 field-strength tensor  is the only physical quantity.
So that, from this point of view,
a polarization-dependent theory in 5D is equivalent to
polarization-independent gravity in 4D coupled to a $U(1)$ field strength.

The $2n$-point functions are further simplified by restricting to a
kinematic region of
  ``high momentum''; meaning that
 we intend to  take only the terms with the highest power of
$\;\omega$, $\;k\equiv|\vec{k}|$ in a given $2n$-point function.
Here, we have introduced the
convention $\;h_{ab}\propto \phi(r)e^{i(\omega t-\vec{k}\cdot\vec{x})}\;$.
This high-momentum region still falls  within the hydrodynamic paradigm, where
the frequency $\omega$ and transverse momentum $k$
are considered to be parametrically lower than the temperature \cite{PSS,SS}.

A subtle point is that, even for this  high-momentum regime, radial derivatives, whether  acting  on  gravitons
or the background, cannot  be immediately disregarded.
This is because $\;g^{rr}\nabla_r\nabla_r\sim \frac{r^2}{r^2}\sim 1\;$, whereas ({\it e.g.})
$\;g^{tt}\nabla_t\nabla_t \sim  \frac {\omega^2}{r^2}\propto r^{-2}\;$.
Hence, the radial derivatives seem to dominate at the AdS boundary for any finite values of  $\omega$ and $k$.
However, the process of  holographic renormalization
\cite {dBVV,skenderis1,skenderis2} for
bulk quantities requires negative powers of $r$ \cite{BwithG} to survive. Then, since our ultimate interest is the gauge theory, a derivative will always implicitly mean either $\nabla_t$ or
$\;\nabla_z\equiv\vec{k}\cdot\vec{\nabla}\;$.

\subsection{Expanding the Lagrangian}

Let us close this section with some useful comments about
perturbatively expanding the metric and our general strategy for expanding $\sqrt{-g}{\cal L}$. We adopt  the 't Hooft--Veltman \cite{tHV} convention, whereby the expansion of any covariant metric (or metric with both indices down) stops at linear order. That is,
\be
g_{ab}\;=\;{\overline g}_{ab}+h_{ab}
\ee
is exact to all orders.  Note that
an overlined quantity signifies the background and indices on a graviton
are always raised by a background (contravariant)  metric.

One then finds that
\bea
g^{ab}\;&=&\;{\overline g}^{ab}-h^{ab}+h^{a}_{\ c}h^{cb}+{\cal O}[h^3]\;,
\label{contra}
\\
\sqrt{-g}\;&=&\;\sqrt{-{\overline g}}\left[1+\frac{1}{2}h^a_{\ a}
-\frac{1}{4}h^a_{\ c}h^c_{\ a} +\frac{1}{8}(h^a_{\ a})^2\right]
+{\cal O}[h^3]\;.
\label{det}
\eea

When carrying out the calculations, we arrange that calculation so that only covariant gravitons are acted on by derivatives. Then, since there can be at most one derivative per graviton (see Appendix~\ref{argue}),
the following exact expression  suffices
to handle  all appearances of a differentiated graviton:
\be
\Gamma_{abc}\;=\;{\overline \Gamma}_{abc}+\frac{1}{2}
\left[{\overline\nabla}_a h_{bc}
+{\overline\nabla}_b h_{ac}
-{\overline\nabla}_ch_{ab}\right]\;.
\label{gamm}
\ee

The following  second-order  expansion also  proves
to be useful:
\be
\delta{\cal R}_{abcd}[h^2]\;=\;\left[\nabla_c-{\overline\nabla}_c\right]
\Gamma_{bda}(h)
\;-\;\Big\{c\leftrightarrow d\Big\}\;,
\label{del3}
\ee
where we have used $\;\nabla-{\overline\nabla} \sim \Gamma(h)\;$.

What is left is to expand out of the contravariant metrics and the determinant. As these gravitons are undifferentiated, they must be tensors. So, the task  simplifies. The relevant expressions
are now
\bea
g^{xx}\;&=&\;{\overline g}^{xx}+h^{x}_{\ y}h^{yx}
+\left(h^{x}_{\ y}h^{yx}\right)^2+\dots+\left(h^{x}_{\ y}h^{yx}\right)^{p}+\dots\;,
\label{contra2}
\\
\sqrt{-g}\;&=&\;\sqrt{-{\overline g}}\Biggl[1
-\frac{1}{2}h^x_{\ y}h^y_{\ x} -\frac{1}{4\cdot2!}\left(h^{x}_{\ y}h^{yx}\right)^2
-\frac{3}{2^3\cdot 3!}\left(h^{x}_{\ y}h^{yx}\right)^3
\dots \nonumber \\
&&\hspace{.7in} -\;\Theta(p)
\left(h^{x}_{\ y}h^{yx}\right)^p\;-\dots\Biggr]\;,
\label{det2}
\eea
such that
\be
\Theta(p)\;\equiv\; \frac{\Gamma\left[p-\frac{1}{2}\right]}
{2\sqrt{\pi} p!}\;,\;\;\;\;\;\;p\;=\;0,1,2,\dots\;.
\label{theta}
\ee

In Eqs.~(\ref{contra2}-\ref{theta}) we have made the physically motivated choice of  $D=5$, which is the case from now on. None of our conclusions would change for larger values of $D$.

\section{The Einstein n-point functions}

We first recall the Einstein
Lagrangian
$\;{\cal L}_E=(1/16\pi G_5){\cal R}\;$ and its variation with respect to the
Riemann tensor,
\be
{\cal X}_E^{abcd}\;
=\;\frac{1}{32\pi G_5}\left[g^{ac}g^{bd}-g^{ad}g^{bc}\right]\;,
\label{XE}
\ee
where $G_5$ is the five-dimensional Newton's constant.

\subsection{Two-point functions}

Let us  begin here with  the two-point functions.
Because of the high-momentum restriction,
we only take into account terms in which every available derivative
acts on a graviton.
As the scalar modes have been deemed irrelevant,
the only possibilities are two differentiated tensor modes or two
differentiated vector modes. A ``mixed combination'' of a tensor and a vector cannot contribute since general covariance requires any term to have
an even number of both $x$ and $y$ indices, and
$\;\nabla_x$=$\nabla_y=0\;$.

By way of  Eqs.~(\ref{gamm},\ref{del3},\ref{XE}) and
some simplification, the case
of two tensor modes work can be worked out.
Using the
notation $\;h_{ab}^{(j)}\propto \exp\left[i\omega_jt-k_jz\right]\;$,
we find
\be
\langle h_2h_2 \rangle_E\;=\; -\frac{1}{32\pi G_5}\sqrt{-g}g^{xx}g^{yy}
\left[h_{xy}^{(1)}\left(\omega_1 g^{tt}\omega_2
\;+\;k_1 g^{zz}k_2\right)h^{(2)}_{xy}\right]\;.
\ee
Here and throughout, the large-momentum regime is implied.

When there are, rather, two vector modes, the result is then
\bea
\langle h_1h_1 \rangle_E\;&=&\; \frac{1}{16\pi G_5}\sqrt{-g}g^{xx}
g^{zz}\left(-g^{tt}\right)
\left[\omega_1h_{zx}^{(1)}+k_1h^{(1)}_{tx}\right]
\left[\omega_2h_{zx}^{(2)}+k_2h^{(2)}_{tx}\right] \cr
&=&\frac{1}{16\pi G_5}\sqrt{-g}g^{xx}
g^{zz}\left(-g^{tt}\right) F^{(1)}_{tz} F^{(2)}_{tz}\;.
\eea
All expressions should be understood  as symmetrized
with respect to  $x$ and $y$, so that ({\it e.g.})
$g^{xx}h_{zx}h_{tx}$ really means $\frac{1}{2}\left[g^{xx}h_{zx}h_{tx}+
g^{yy}h_{zy}h_{ty}\right]$.

\subsection{Higher-point functions}

Because vector modes appear only through $F$ and so must be differentiated,
 and because undifferentiated tensor modes
can only be added in pairs ({\it cf}, Eqs.~(\ref{contra2},\ref{det2})),
the $n$-point functions with odd numbers of gravitons vanish. So let us next consider the $2n$-point functions with $n\ge 2$.
These could either be worked out by a  brute force expansion or deduced from the two-point functions by  way of  simple combinatorial arguments.

The coefficients of the contravariant-metric expansion are given in Eq.~(\ref{contra2}) and those of the determinant, in Eq.~(\ref{det2}),
which leads us to
\bea
\langle (h_2)^{2n}\rangle_E\;&=&\; \dbinom{2n}{2}\sum_{p=0}^{n-1}(n-p)\Theta(p) \frac{\sqrt{-g}g^{xx}g^{yy}}
{32\pi G_5} \left[\prod_{j=2}^{n} (h^x_{\ y})^{(2j-1)} (h^y_{\ x})^{(2j)}\right]
\nonumber  \\
&\times&\left[h_{xy}^{(1)}\left(\omega_1 g^{tt}\omega_2
\;+\;k_1 g^{zz}k_2\right)h^{(2)}_{xy}\right]\;.
\eea
In the previous equation, the binomial  factor in front of the sum accounts for the number of ways of drawing two (differentiated) tensor modes out of the $2n$ available,
the summation index counts  the number of pairs
of modes in the expansion of the determinant and  the factor of  $\;(n-p)\;$ is the number of ways of drawing
the remaining $\;n-1-p\;$ pairs out of two contravariant
metrics. Here, we have used that the number of ways of  drawing
$q$ identical objects from $m$ distinct ``boxes'' is $\dbinom{q+m-1}{m-1}$.
The summation can be done explicitly,
\bea
\label{2npointI}
\langle(h_2)^{2n}\rangle_E\;&=&\;-\dbinom{2n}{2} \frac{\Gamma\left[n+\frac{1}{2}\right]}{\sqrt{\pi}
\Gamma[n]}\frac{\sqrt{-g}g^{xx}g^{yy}} {16\pi G_5}
\left[\prod_{j=2}^{n}(h^x_{\ y})^{(2j-1)}(h^y_{\ x})^{(2j)}\right] \nonumber \cr
&\times&\left[h_{xy}^{(1)}\left(\omega_1 g^{tt}\omega_2
\;+\;k_1 g^{zz}k_2\right)h^{(2)}_{xy}\right]\;\nonumber \cr
&=&\;- \frac{(2n-1) \Gamma\left[n+\frac{1}{2}\right]}{\sqrt{\pi}
\Gamma[n-1]}\frac{\sqrt{-g}g^{xx}g^{yy}} {16\pi G_5}
\left[\prod_{j=2}^{n}(h^x_{\ y})^{(2j-1)}(h^y_{\ x})^{(2j)}\right] \\
&\times&\left[h_{xy}^{(1)}\left(\omega_1 g^{tt}\omega_2
\;+\;k_1 g^{zz}k_2\right)h^{(2)}_{xy}\right]\; .
\eea

For the case of two vector
modes, there is no need for a leading binomial factor
and only one contravariant metric is expanded; and similar methods yield
\bea
\label{2npointII}
\langle (h_1)^2 (h_2)^{2n-2}\rangle_E\;&=&\; \sum_{p=0}^{n-1}\Theta(p)\frac{\sqrt{-g}g^{xx}g^{zz}
g^{tt}}
{16\pi G_5}
F_{tz}^{(1)}F_{tz}^{(2)}
\prod_{j=2}^{n}(h^x_{\ y})^{(2j-1)}(h^y_{\ x})^{(2j)}
\nonumber
\\
&=&\; -\frac{\Gamma\left[n-\frac{1}{2}\right]}{\sqrt{\pi}
\Gamma[n]} \frac{\sqrt{-g}g^{xx}g^{zz}
g^{tt}}
{16\pi G_5}
F_{tz}^{(1)}F_{tz}^{(2)}
\prod_{j=2}^{n}(h^x_{\ y})^{(2j-1)}(h^y_{\ x})^{(2j)}\!. \hspace{.3in}
\eea
Equations~(\ref{2npointI}) and (\ref{2npointII}) exhaust all possible $n$-point functions.

\section{The Gauss--Bonnet  n-point functions}

\subsection{Initial considerations}

We view the various Gauss--Bonnet expressions as extensions to the leading Einstein term. So, the Lagrangian for this theory is
(with $L$ momentarily restored)
\be
\frac{1}{16\pi G_5}{\cal R} +  \frac{1}{L}{\cal L}_{GB}
\;=\; \frac{1}{G_5}\left[\frac{\cal R}{16\pi} +
\frac{l_p^2}{L^2}  L^2{\cal L}_{GB}\right]\;,
\ee
where we have used  $\;l_p=\sqrt{G_4}\sim\sqrt{G_5/L}\;$.
This makes it clear that the relative
strength of the Gauss-Bonnet extension  goes as
$l_p^2/L^2$, which is parametrically smaller than unity.

Let us now recall the Gauss--Bonnet Lagrangian and its variation,
\bea
{\cal L}_{GB}\;&=&\;\lambda\left[{\cal R}^{abcd}R_{abcd}
-4{\cal R}^{ab}{\cal R}_{ab}+{\cal R}^2\right]\;,
\\
{\cal X}_{GB}^{abcd}\;&=&\;\lambda\bigl[{\cal R}^{abcd}
-{\cal R}^{abdc}- 2g^{ac}{\cal R}^{bd} -2g^{bd}{\cal R}^{ac} 
\nonumber \\
&& \;+\;2g^{ad}{\cal R}^{bc} +2g^{bc}{\cal R}^{ad} + {\cal R}g^{ac}g^{bd}
-{\cal R}g^{ad}g^{bc}\bigr]\;,
\eea
where $\lambda$ is  a  dimensionless number of order
unity.

We will first look at the  four-point functions.
In light of previous considerations, there are only three viable ways of selecting the four  gravitons:
four tensor modes, four vector modes or  two of each.
We will, however, proceed to argue that only the first of these choices can have any physical relevance and, even for this one, the calculation is much simpler than it might appear.

\subsection{Simplifying the Gauss--Bonnet calculations}

Let us begin here with the case of four tensor modes.
It is only necessary to include the contribution from the Riemann-tensor-squared term of ${\cal L}_{GB}$ since it already contains all physical information about the scattering of four tensor modes \cite{Zwei} (and references therein). This claim can  be
readily  understood from the perspective of  field redefinitions \cite{tH,POL}.

To clarify the above argument, suppose that  we start with  the following term in the two-point function,
$h_{ab}h^{ab}$. Now, redefine the tensor modes  $\;h_{ab}\to h_{ab}+ \delta^{(1)}{\cal R}_{ab}+ \delta^{(2)}{\cal R}_{ab}+\cdots\;$.
One of the products of this transformation goes as
$\;\delta^{(2)}{\cal R}^{ab}\delta^{(2)}{\cal R}_{ab}\;$;
that is, precisely the fourth-order contribution from Ricci-tensor-squared term. Similarly, we can reproduce the fourth-order contribution from the Ricci-scalar-squared term with the redefinition $\;h_{ab}\to h_{ab}+ {\overline g}_{ab}
\delta^{(1)}{\cal R}+{\overline g}_{ab}\delta^{(2)}{\cal R}+\cdots\;$. We are, of course, free to combine these (retaining only the relevant parts):
$\;h_{ab}\to h_{ab} + a_1\delta^{(2)}{\cal R}_{ab}+a_2{\overline g}_{ab}\delta^{(2)}
{\cal R}\;$.
Now, if one wants to do away with the fourth-order contributions
from the Ricci-tensor-squared and the Ricci-scalar-squared terms, it becomes the matter of appropriately  choosing the numerical coefficients $a_1$ and $a_2$.

There is, however, no such field redefinition  that
can produce the fourth-order term from the  Riemann-tensor-squared term.
Each graviton has two symmetric indices,  and so  a contraction like $h_{ab}h^{ab}$ cannot reproduce the requisite four-index structure of the Riemann-tensor-squared expansion.

To sum up, of the three  Gauss--Bonnet terms, only the  Riemann-tensor-squared term is  of  physical relevance.

By similar reasoning, one can  argue that {\em any} (four-derivative) four-point function with
vector modes is devoid of physical meaning, irrespective of the interactions. This is because, as previously discussed, vector interactions can be represented in terms of field-strength
tensors ({\it cf}, Eq.~(\ref{FST})) which do not involve
$x$ or $y$ indices. As a consequence, a  fourth-order term containing vector modes must be one of the four simple forms ---
$\;F_{ab}F_{cd}F^{ad}F^{bc}\;$,
$\;F_{ab}F^{ab}F_{cd}F^{cd}\;$,
$\;F_{ab}F^{ab}\nabla_ch_{de}\nabla^ch^{de}\;$,
or
$\;F_{ac}F^{c}_{\ b}\nabla^ah_{de}\nabla^bh^{de}\;$ ---
any of which can be  attained by suitably redefining a graviton.
For instance,  $\langle FF\rangle$ and  $\;F_{ab}\to
F_{ab}+F_{ac}F^{c}_{\ b}\;$ leads to the first form,
$\langle  FF\rangle$ and  $\;{\overline g}_{ab}\to
{\overline g}_{ab} + h_{ab} \to {\overline g}_{ab}+h_{ab}+
F_{ac}F^{c}_{\ b}\;$
yields the second (via the determinant); whereas  $\langle h_2h_2\rangle$ and the preceding transformation gives us the latter pair (respectively by way of  the determinant and a contravariant metric).

There is yet  another argument  that allows us to reach the same conclusion about the  vector-mode amplitudes. Gauss--Bonnet gravity leads to equations of motion that are at most quadratic in derivatives. So a fourth-order expansion of its Lagrangian in what are  field-strength tensors  had better give back either the
fourth-order term in the  Born--Infeld Lagrangian, since Born--Infeld's theory
\cite{BI} is the electromagnetic analogue of Lovelock gravity \cite{BTZ}, or nothing at all. Our actual calculations of  $\langle h_1h_1h_1h_1\rangle_{GB}$
do indeed lead to the latter result. Meanwhile, the ``mixed'' four-point function  $\langle h_1h_1h_2h_2\rangle_{GB}$ is
constrained (and found) to vanish by  similar reasoning, as the possible form
of term in the Lagrangian producing such  an amplitude is
${\cal R}_{abcd}F^{ad}F^{bc}$, and this would lead to equations of motion with higher than two derivatives.

\subsection{The results}

And so the four-point functions amount to  a single calculation,
expanding Riemann-tensor-squared to fourth order in tensors.  We have performed this expansion and obtained
\bea
\langle h_2h_2h_2h_2\rangle_{GB}\;&=&\; \frac{3}{4}\lambda\sqrt{-g}(g^{xx})^2(g^{yy})^2
\left[h_{xy}^{(1)}\left(\omega_1 g^{tt}\omega_2
\;+\;k_1 g^{zz}k_2\right)h^{(2)}_{xy}\right]\nonumber
\\
\;\;\;\;\;\;\;\;\;\;\;\;\;\;\;&\times&
\left[h_{xy}^{(3)}\left(\omega_3 g^{tt}\omega_4
\;+\;k_3 g^{zz}k_4\right)h^{(4)}_{xy}\right]\;.
\eea
To keep the calculation tractable, it is better
to keep all linearized $\Gamma$'s in their covariant form
as in Eq.~(\ref{gamm}) and apply the Riemann (anti-) symmetry properties only at the end of the calculation.

Let us now move on to  the $2n$-point functions with $n\geq 4$.
We need, of course, only consider the prospect of having all tensors, as adding  additional pairs of undifferentiated gravitons cannot invalidate the previous
arguments.

Again calling upon simple combinatorics, we find that
\bea
\langle (h_2)^{2n} \rangle_{GB}\;&=& \;-\frac{3}{4}\lambda\dbinom{2n}{4}\sum_{p=0}^{n-2}
\dbinom{n-p+1}{3}\Theta(p)
\sqrt{-g}(g^{xx}g^{yy})^2
\nonumber
\\
&\times&
\left[\prod_{j=2}^{n}(h^x_{\ y})^{(2j-1)}(h^y_{\ x})^{(2j)}\right]
\prod_{l=1}^{2}\left[ h_{xy}^{(2l-1)}\left(\omega_{2l-1} g^{tt}
\omega_{2l}
\;+\;k_{2l-1} g^{zz}k_{2l}\right)h^{(2l)}_{xy}\right]
\nonumber \\
 &=&\frac{2}{5}\lambda\dbinom{2n}{4} \frac{\Gamma\left[n+\frac{3}{2}\right]}{\sqrt{\pi}
\Gamma[n-1]} \sqrt{-g}(g^{xx}g^{yy})^2 \times \left[\prod_{j=2}^{n}(h^x_{\ y})^{(2j-1)}(h^y_{\ x})^{(2j)}\right]
 \nonumber \\
&\times&
\prod_{l=1}^{2}\left[ h_{xy}^{(2l-1)}\left(\omega_{2l-1} g^{tt}
\omega_{2l}
\;+\;k_{2l-1} g^{zz}k_{2l}\right)h^{(2l)}_{xy}\right]\;.
\eea
In the top line, the left-most binomial  factor is the number
of ways of drawing  four differentiated tensors modes from the $2n$ available,
the summation index  is again counting the  pairs
of modes that are drawn out of the  determinant and  the
right-most binomial factor accounts for the number of  ways of extracting
the $\;n-2-p\;$ remaining  pairs from the four contravariant
metrics.

This exhausts the possible $n$-point functions in the high-momentum regime.

\section{Comparing Einstein and Gauss-Bonnet}

\subsection{Angular dependence of the $n$-point functions}

We can use the results of Sections~3 and~4, to express the statements about scattering angles in a precise way.
We work at $\;r\to\infty\;$, as appropriate for making contact with the gauge theory, although a different choice of $r$ would be
inconsequential.~\footnote{An exception is at the brane horizon,
where the divergence
of $g^{tt}$ effectively wipes out all information about the transverse
momenta. So that, for a hypothetical scattering experiment
on the horizon, the $n$-point functions would appear angular independent
for any theory and for  any $n$.}

Let us begin with
the Einstein $2n$-point functions and assume, for the moment,
only tensor modes. Then, for $n=1$,
\be
\lim_{r\to\infty} \langle h_2h_2 \rangle_{E} \;\sim\;  h^{(1)}_{xy}\left[\omega_1\omega_2 -
\vec{k}_1\cdot \vec{k}_2
\right]h^{(2)}_{xy}\;,
\ee
where the arbitrariness of the propagation direction  has now been made explicit and the $\sim$ indicates some normalization factors that are not essential to our discussion. Here, it becomes the simple matter of applying momentum conservation.
That is, $\vec{k_2}=-\vec{k_1}$,  and so we obtain the
angular-independent form
\be
\lim_{r\to\infty} \langle h_2h_2\rangle_{E} \;\sim\;  k_1^2 \;   h^{(1)}_{xy}    h^{(2)}_{xy}\;.
\ee
Although not yet crucial, the on-shell condition $\;\omega_{j}=k_j\equiv|\vec{k}_{j}|\;$ has also
been imposed.

Continuing to larger values of $n$,
we find that
\bea
\label{final2n}
\lim_{r\to\infty}\langle h_2\dots h_2 \rangle_{E} &\sim& h^{(1)}_{xy}\left[k_1k_2\left(1-\cos\theta
\right)\right]
h^{(2)}_{xy}\; \nonumber \\
&\sim&
 s\; h^{(1)}_{xy} h^{(2)}_{xy}\;,
\eea
for any even number of gravitons greater than two.
In the second equality of Eq.~(\ref{final2n}), we have introduced the Mandelstam variable $\;s=k_1^\mu k_{2\mu}\;$ such that $\;k_i^{\mu}=(\omega_i,\vec{k}_i)\;$. Recall that, for massless particles, the sum of the three Mandelstam variables vanishes, $\;s+t+u=0\;$ ($\;t=-k_1^\mu k_{3\mu}\;$, $\;u=-k_1^\mu k_{4\mu}\;$).

The situation is even  simpler for a $2n$-point function that
contains two necessarily differentiated vector modes.
One can deduce that there is never any angular dependence
for any value of $n$ by simply recognizing that the  explicit frequencies and momenta are already included in the definition of the  physical modes $F^{(1,2)}_{tz}$ and there are no other derivatives available to introduce  additional
angular dependence.

Let us now find the corresponding $n$-point functions for the Gauss--Bonnet
theory, as these can then be compared to the Einstein expressions. Given our interest in the high-momentum regime,  the simplest case is the four-point
function. When symmetrized with respect to the four gravitons, this goes as
\bea
\lim_{r\to\infty}\langle h_2h_2h_2h_2\rangle_{GB} &\sim&  h^{(1)}_{xy}\left[\omega_{\left(1\right.}\omega_2 -
\vec{k}_{\left(1\right.}\cdot \vec{k}_2
\right]h^{(2)}_{xy}
h^{(3)}_{xy}\left[\omega_3\omega_{\left.4\right)} -
\vec{k}_3\cdot \vec{k}_{\left.4\right)}
\right]h^{(4)}_{xy}
\;
\nonumber \\
 &\sim&  h^{(1)}_{xy}\left[
k^{\mu}_{\left(1\right.}\cdot k_{2\mu}
\right]h^{(2)}_{xy}
h^{(3)}_{xy}\left[
k_{\mu 3}\cdot k^{\mu}_{\left.4\right)}
\right]h^{(4)}_{xy}\;,
\eea
which can, when on-shell, be simplified in terms of the Mandelstam variables,
\bea
\lim_{r\to\infty}\langle h_2h_2h_2h_2\rangle_{GB} &\sim& h^{(1)}_{xy}h^{(2)}_{xy}
\left[-s(t+u)\right]h^{(3)}_{xy}
h^{(4)}_{xy} \nonumber \\
&\sim&  s^2\; h^{(1)}_{xy}h^{(2)}_{xy} h^{(3)}_{xy}
h^{(4)}_{xy}
\;.
\label{angle2}
\eea

That leaves us to look at the Gauss--Bonnet $2n$-point functions
with $n\ge3$. As for the analogous Einstein calculation,
the condition of momentum conservation
$\;\sum_{j=1}^{2n}\vec{k}_{j}=0\;$
is no longer useful; meaning that two angles now require
specification.
A simple way to account for  this new angle is to introduce
a ``generalized Mandelstam variable''
$\;v = -k_1^\mu\sum_{j=5}^{2n} k_{j\mu}\;$,
for which it is readily confirmed that
$\;s+t+u=-v\;$.
Then, with the on-shell condition imposed,
\bea
\lim_{r\to\infty}\langle h_2h_2h_2h_2\rangle_{GB} &\sim& h^{(1)}_{xy}h^{(2)}_{xy}
\left[-s(t+u)\right]h^{(3)}_{xy}
h^{(4)}_{xy} \nonumber \\
&\sim&  s(s+v)\; h^{(1)}_{xy}h^{(2)}_{xy} h^{(3)}_{xy}
h^{(4)}_{xy}
\;.
\eea

\subsection{The gauge-theory perspective}

So as to connect with experiment, our ultimate interest is in the corresponding
$n$-point  stress-tensor correlators for the gauge-theory dual.
To this end, the standard prescription is to send a  bulk quantity toward the AdS boundary and then apply standard subtraction techniques before taking the final $\;r\to\infty\;$ limit
\cite{dBVV,skenderis1,skenderis2}.  And so it should, with some  effort, be possible to translate our results into statements about the gauge theory \cite{future}.

The stress-energy tensor correlators of the gauge theory can be expected to inherit the angular dependence of the graviton $n$-point functions. This is because the subtraction process is equivalent to a process of  matching and then stripping off the divergent bulk and boundary conformal factors
\cite{Witten,skenderis0,stuff,things}. Such a process would not change the angular dependence of the correlators because the  metric components $g^{tt}$ and
$g^{zz}$ are dispersed democratically and exhibit the same radial
dependence at the boundary.

\section{Conclusion}

Beginning with the premise of a UV-complete gauge theory
and its UV-complete gravitational dual,
we applied our argument \cite{prepap} that an effective theory describing  gravitational perturbations about a background solution  must be organized into Einstein gravity plus terms of the Lovelock class.  The leading-order effective description must then either be Einstein gravity or a Gauss--Bonnet extension thereof.

Given an AdS black brane background solution and
a kinematic regime of high-momentum, we
have calculated, for any number of gravitons,
all of the physical $n$-point functions. This
was done for both of the proposed effective theories, with  all the results
having
been  expressed in terms
of gauge-invariant gravitational modes.

We have, from a novel perspective, explained why the Einstein
$n$-point functions  have a simpler angular dependence
than those of Gauss--Bonnet gravity and then used our results to quantify the angular dependence of both theories in a precise manner.

The graviton
$n$-point functions have a direct correspondence with
stress-energy tensor correlators in the gauge theory.
Holography implies that these gauge-theory
correlators should inherit the same angular dependence.
This means that there should be fundamental and testable distinction between the Einstein and Gauss--Bonnet models.
Following  ideas from  \cite{Mald-Hof,Hof}, we have proposed
that heavy-ion scattering experiments can be used for such purposes.

\section*{Acknowledgments}

The research of RB was supported by the Israel Science Foundation grant no. 239/10. The research of AJMM received support from Rhodes University and the Claude Leon Merit Award grant no.
CLMA05/2011. AJMM thanks the Arnold Sommerfeld Center for Theoretical Physics at Ludwig-Maximilians-Universitat Munchen for their hospitality.

\appendix

\section{The case against two derivatives on a graviton} \label{argue}

In working within the high-momentum regime, we require that all
derivatives (two for Einstein, four for Gauss--Bonnet, {\it etc.}) act on a graviton. At a first glance, it would appear that there could well be terms with two derivatives acting on the same graviton. However, as we now show, such terms cannot contribute on-shell. This can be verified by explicit calculations
but can also be understood through the following simple argument, whose domain of applicability is discussed at the end.

Let us demonstrate that the above claim is true for any Lovelock
theory with four derivatives acting on gravitons. The
extension to other cases, including the simplest case of two derivatives, is then  straightforward.

The linearized field equation for any
Lovelock theory can be expressed as
\be
{\overline {\cal X}}^{abcd} \delta^{(1)}{\cal R}_{abcd}\;=\;0\;,
\label{LFE}
\ee
such that
\be
\delta^{(1)}{\cal R}_{abcd}\;=\;{\overline\nabla}_c\Gamma_{bda}(h)-
{\overline\nabla}_d\Gamma_{bca}
(h)\;.
\label{del1}
\ee
Here,  a numerical superscript  denotes the  number of
gravitons,
 an over-lined quantity signifies the background and, in this discussion,
 we often
neglect the usual (anti-) symmetrization
of indices.

The simplicity of the field equation~(\ref{LFE})
follows from that of  a  generic theory of
gravity
\cite{Wald,BGHM},
\be
2\nabla_b\nabla_a {\cal X}^{apqb} \;-\;
{\cal X}^{abcp}{\cal R}_{abc}^{\;\;\;\;\;\;\;q}
\; +\; \frac{1}{2}g^{pq}{\cal L}
\;=\;0\;,
\label{fieldeq}
\ee
along with the Lovelock
identity~(\ref{LLidentity}).~\footnote{For the relevant modes (vectors and
tensors),
the third term in the
field equation must be a mass
term and ends up being absorbed into
the mass terms of
 $\;\delta^{(1)}{\cal R}\sim \nabla\Gamma(h)\sim \Box h\;$.}

Let us next look at  the fourth-order expansion
of the Lagrangian density of a Lovelock theory.
Considering just the terms with exactly four derivatives acting on
gravitons,
we have (up to inconsequential numerical factors)
\be
\frac{\delta^{(4)}\left(\sqrt{-g}{\cal L}\right)}{\sqrt{-{\overline g}}}
\;=\;
\left[h\delta^{(1)}{\cal R}_{pqrs}
+ \delta^{(2)}{\cal R}_{pqrs}\right]
{\overline {\cal Y}}^{\substack{abcd\\pqrs}}
\left[h\delta^{(1)}{\cal R}_{abcd}
+ \delta^{(2)}{\cal R}_{abcd}\right]\;,
\ee
where  ${\cal Y}$ is the variation of ${\cal X}$ with respect to
${\cal R}$ (indices suppressed) and
a depicted ``$h$'' is  meant to indicate that a
single undifferentiated  graviton from the expansion of the determinant or the  contravariant metric. Notice that $\;\delta^{(2)}{\cal R}\sim \Gamma(h)
\Gamma(h)\;$; {\it cf}, Eq.~(\ref{del3}).

The crucial point is that the tensor ${\cal Y}$
inherits, just like ${\cal X}$ does, all the (anti-) symmetry
properties of the Riemann tensor. In fact, this tensor has two sets
of four indices, each of which is Riemannian in structure.
Hence, its contraction with $\;h\delta^{(1)}{\cal R}\;$,
either from the left or
the right, must necessarily  give back  a form that
is proportional to the linearized field equation (however, see below).
Meaning that
the on-shell form of this fourth-order
Lagrangian density is simply
\be
\frac{\delta^{(4)}\left(\sqrt{-g}{\cal L}\right)}{\sqrt{-{\overline g}}}
\;=\;
\delta^{(2)}{\cal R}_{pqrs}
\left({\overline {\cal Y}}^{\substack{abcd\\pqrs}}\right)
\delta^{(2)}{\cal R}_{abcd}\;.
\ee

The same basic argument persists
for any number of pairs of derivatives all acting on gravitons.
That is, it can be applied to any order of Lovelock theory
(and, in particular, Einstein gravity)
with always the same outcome: no more than one derivative per graviton.

Strictly speaking, this argument is only rigorous
 at the AdS boundary
and at the horizon, as these are the only the surfaces where
the metric and its descendants (${\cal R}$, ${\cal X}$, ${\cal Y}$, {\em etc.})
are assured to be insensitive to the polarization.~\footnote{The horizon,
effectively so, because the metric reduces to a 1+1 conformally
flat space.} This being a sufficient (albeit not necessary) condition
for $\;{\cal Y}\cdot \delta^{(2)}{\cal R} \propto {\cal X}\cdot
\delta^{(2)}{\cal R}\;$.
However, at any radius, the background metric can be
regarded as (polarization-independent) Einstein
 plus ${\cal O}[l_p^2/L^2]$ corrections.
Hence, any violation of this argument is suppressed by an {\em additional}
factor of $l_p^2/L^2$ relative to other contributions
from the same Lagrangian.

\end{document}